\documentclass[a4paper,english,aps,floats,twocolumn,showpacs,nofootinbib]{revtex4}
\usepackage{pslatex}
\usepackage[T1]{fontenc}
\usepackage{graphicx}
\usepackage{epsfig}

\usepackage{calc}
\usepackage{ifthen}
\usepackage{subfigure}

\newcommand{\bea}{\begin{eqnarray}}
\newcommand{\eea}{\end{eqnarray}}

\newcommand{\nc}{\newcommand}
\nc{\renc}{\renewcommand}
\nc{\eqs}[2]{\mbox{Eqs.~(\ref{#1},\,\ref{#2})}}
\nc{\eq}[1]{\mbox{Eq.~(\ref{#1})}}
\nc{\figs}[2]{\mbox{Figs.~(\ref{#1},\,\ref{#2})}}
\nc{\fig}[1]{\mbox{Fig~.(\ref{#1})}}
\nc{\be}[1]{\begin{equation} \mbox{$\label{#1}$}}
\nc{\ee}{\vspace{0.1cm}\end{equation}}

\newcommand{\bean}{\begin{eqnarray*}}
\newcommand{\eean}{\end{eqnarray*}}

\def\MeV{{\rm \ MeV}}

\def\lae{\;^{<}_{\sim} \;}

\nc{\npp}[3]{{\it  Nucl.\ Phys.\ }{{\bf #1} {(#2)} {#3}}}
\nc{\prdd}[3]{{\it  Phys.\ Rev.\ D\ }{{\bf #1} {(#2)} {#3}}}
\nc{\prll}[3]{{\it Phys.\ Rev.\ Lett.\ }{{\bf #1} {(#2)} {#3}}}
\nc{\pll}[3]{{\it  Phys.\ Lett.\ }{{\bf #1} {(#2)} {#3}}}

\begin{document}
\title{Comment on the "Freeze-In" mechanism of dark matter production 
}
\author{John McDonald}
\email{j.mcdonald@lancaster.ac.uk}
\affiliation{Cosmology and Astroparticle Physics Group, Dept. of Physics, University of
Lancaster, Lancaster LA1 4YB, UK}
\begin{abstract}

    We point out an earlier implementation of the "freeze-in" mechanism for the production of very weakly-interacting dark matter. We also clarify a footnote given in 0911.1120  regarding this eariler paper.

\end{abstract}
\maketitle

\section{Introduction}

         In \cite{fimpc} a mechanism for production of very weakly-interacting dark matter was proposed. The idea is that particles which are in thermal equilibrum can decay to stable particles which are very weakly-coupled to the thermal equilibrium particles. Being very weakly-coupled, the stable particles remain out of thermal equilibrium and become a density of dark matter. The majority of dark matter particles are produced by decays when $T \approx M$, where $M$ is the mass of the thermal equilibrium particle. Thereafter the number density of the decaying particles becomes Boltzmann-suppressed, effectively switching off dark matter production.  

      This mechanism was called "freeze-in" in \cite{fimpc}. The purpose of this short note is to point out that \cite{fimpc} was not the first paper to introduce and apply this mechanism. It had been introduced several years earlier in \cite{therm}.  

      In \cite{therm}, exactly this mechanism was applied to the specific case of the 
production of gauge singlet scalar dark matter by decay of Higgs bosons via the 'Higgs portal' coupling.  The results of \cite{therm} can be summarised as follows. 
The portal coupling of gauge singlets $S$ to the Higgs doublet $H$, $\lambda S^{\dagger}SH^{\dagger}H$, results in a linear coupling to the physical Higgs boson $h$ once the Higgs VEV is introduced. This allows the Higgs bosons in the thermal bath to decay to pairs of singlets. 
An essential requirement for dark matter production by Higgs decays is that the singlets $S$ remain out of thermal equilibrium. As first noted in \cite{bento}, this requires that the coupling $\lambda$ is very weak, $\lambda \lae 10^{-10}$. 
In this case the density of $S$ particles produced in Higgs decay will freely expand to become a density of dark matter. In \cite{therm} the resulting density was calculated by integrating the thermal distribution of Higgs bosons down from a high $T$. It was found that the largest contribution to 
the present dark matter density comes from Higgs decays when $T \approx m_{h}$, after which the 
Higgs bosons become Boltzmann suppressed. (Annihilation of Higgs boson in the thermal bath can also contribute to the dark matter density, but in was shown in \cite{therm} that this is typically a sub-leading effect.)  It was found that the correct density of $S$ dark matter is produced when $\lambda$ is not very much less than the thermal equilibrium upper bound, $\lambda 
\approx 10^{-10}$.

      Gauge singlet dark matter is generally a two parameter model, corresponding to the mass term $m$ and the coupling $\lambda$. In \cite{therm} it was noted that the physical $S$ mass could be entirely produced by the Higgs VEV, in the limit where $m$ is very small. In that case the model 
becomes a one parameter model, which may be chosen to be the $S$ mass. Therefore the $S$ mass is fixed by the $S$ dark matter density. This turns out to be $\approx 3 \MeV$. 

    \cite{therm} was motivated by the possibility that such very weakly interacting scalars could serve as self-interacting dark matter (SIDM). This happens to require that the $S$ mass is in the range 1-10 MeV for $S$ self-coupling $\sim 0.1$, an intriguing coincidence. 
  
    The existence and content of \cite{therm} was communicated to the authors of \cite{fimpc} shortly after \cite{fimpc} appeared on the arXiv. However, the subsequent citation of \cite{therm} in 
\cite{fimpc}, in a footnote, states that "The phenomenology of this interaction has been previously considered in ref.[23].". This is not an accurate description of the content of \cite{therm}.  In particular, it does not inform the reader that \cite{therm} is in fact an earlier implementation of the "freeze-in" mechanism.  

     We hope that this short note will clarify the footnote given in \cite{fimpc} (which the present author has only recently become aware of), and will inform the reader of this earlier implementation of the "freeze-in" mechanism.

\end{document}